\UseRawInputEncoding
\pdfoutput=1
\documentclass[superscriptaddress,twocolumn,amsmath,amssymb,aps,pra]{revtex4}
\usepackage{mathrsfs}
\usepackage{graphicx}
\usepackage{subfigure}
\usepackage{dcolumn}
\usepackage{bm}
\usepackage{amssymb}
\usepackage{amsmath}
\usepackage{paralist}
\usepackage{color}
\usepackage{txfonts}
\usepackage{ulem}

\newcommand{\Tr}{\mathrm{Tr}}
\newcommand{ \perm}{\mathrm{ perm}}

\newcommand{ \sinc}{\mathrm{sinc}}

\def\be{\begin{equation}}
\def\ee{\end{equation}}
\def\bea{\begin{eqnarray}}
\def\eea{\end{eqnarray}}

\begin{document}

\title{Quantum  dynamics of Gaudin  magnets}

\author{Wen-Bin He}
\affiliation{Beijing Computational Science Research Center, Beijing 100193, China}
\affiliation{State Key Laboratory of Magnetic Resonance and Atomic and Molecular Physics, 
Wuhan Institute of Physics and Mathematics, APM, Chinese Academy of Sciences, Wuhan 430071, China}
\affiliation{The Abdus Salam International Center for Theoretical Physics, Strada Costiera 11, 34151 Trieste, Italy.}

\author{Stefano Chesi}
 \email{stefano.chesi@csrc.ac.cn}
\affiliation{Beijing Computational Science Research Center, Beijing 100193, China}
\affiliation{Department of Physics, Beijing Normal University, Beijing 100875, China}
\affiliation{The Abdus Salam International Center for Theoretical Physics, Strada Costiera 11, 34151 Trieste, Italy.}

\author{H.-Q. Lin}
\affiliation{Beijing Computational Science Research Center, Beijing 100193, China}
\affiliation{Department of Physics, Beijing Normal University, Beijing 100875, China}

\author{Xi-Wen Guan}
\email[]{xwe105@wipm.ac.cn}
\affiliation{State Key Laboratory of Magnetic Resonance and Atomic and Molecular Physics,
Wuhan Institute of Physics and Mathematics, APM,  Chinese Academy of Sciences, Wuhan 430071, China}
\affiliation{NSFC-SPTP Peng Huanwu Center for Fundamental Theory, Xi'an 710127, China}
\affiliation{Department of Theoretical Physics, Research School of Physics and Engineering,
Australian National University, Canberra ACT 0200, Australia}
\date{\today}


\begin{abstract}
Quantum dynamics of many-body systems is a fascinating and significant subject for both theory and experiment. The question of how an isolated many-body system evolves to its steady state after a sudden perturbation or quench still remains challenging. In this paper, using the Bethe ansatz wave function, we study the quantum dynamics of an inhomogeneous Gaudin magnet. We  derive explicit analytical expressions for various local dynamic  quantities with an arbitrary number of flipped bath spins, such as: the spin distribution function, the spin-spin correlation function, and the Loschmidt echo. We also numerically study  the relaxation behavior of these dynamic properties, gaining considerable insight into coherence and entanglement between the central spin and the bath.
In particular, we find that the spin-spin correlations relax to their steady value via a nearly logarithmic scaling, whereas the Loschmidt echo shows an exponential relaxation to its  steady value. Our results advance the understanding of relaxation dynamics and quantum correlations of long-range interacting models of Gaudin type. 
\end{abstract}
\maketitle

Although mature paradigms, like the celebrated Fermi liquid theory \cite{Landau57a,Landau57b,Landau59} and Luttinger liquid theory \cite{Lut1974,Lut1963}, enable us to understand well a wide class of many-body systems in the stationary state, their dynamical evolution is much less understood.
In fact, quantum dynamics of many-body systems always presents itself with formidable challenges, due to the difficulty of deriving eigenfunctions analytically and the exponentially growing complexity of numerics. 
Despite these theoretical difficulties, significant progresses have been made in the last decade, which greatly improves our understanding of dynamical properties of many-body systems. 
For example, the Kibble-Zurek mechanism (KZM) \cite{zurek85} reveals power-law relations between the typical length scale  $\xi$ of  defect domains, the relaxation time $\tau$, and the rate of change of the driving parameter $\lambda$, namely $\xi\sim \lambda^{-\nu}$ and $\tau \sim \lambda^{-z \nu}$, which has been observed in ion Coulomb crystals \cite{NCion}. 
 Thermalization of isolated many-body systems is  also a difficult  topic. The eigenstate thermalization hypothesis (ETH) \cite{deutsch,Srednicki} asserts  that, for a closed systems evolving to the steady state, the diagonal and micro-canonical ensembles are equivalent. 
In recent experiments  with  ultracold atoms \cite{APolkRMP,GuaBL13}, T. Kinoshita et. al showed that Bose gases in one-dimensional traps do not thermalize, even after thousands of collisions \cite{Kinoshita}. 
The ETH can be violated by integrable models \cite{olshaii}, due to the existence of infinitely many conserved charges, or in the presence of many-body localization \cite{mserbyn}.
The study of dynamics is of crucial importance for many-body quantum systems out of equilibrium, as it provides important clues on if and how their equilibrium state can be reached.

Therefore, while thermalization usually concerns the steady-state properties of many-body systems after a long-time evolution, an interesting question is on how the many-body system evolves at intermediate times. 
For a system with a short-range interaction, like the Heisenberg spin chain and the Lieb-Lininger model, light-cone dynamics dominates the relaxation of local observables to steady-state values, such as spin polarization and correlation functions \cite{Bonnes14}. 
However, the relaxation process of local and global quantities to the steady state is less explored in models with long-range interactions. 
In this paper, we use  Gaudin magnets, i.e. the inhomogeneous central spin models, to study the relaxation dynamics of many-body systems. 
Such  magnets with long-range interactions may describe the decoherence of a spin-qubit due to the interaction with surrounding spins, for example an electron interacting with nuclear spins in  quantum dots or defect centers \cite{CoishBaugh2009,WenYang2017}.
Quantum dynamics of the Gaudin model  has been extensively studied with a variety of methods as, for example, exact diagonalization and quantum many-body expansions \cite{WenYang2017,Dloss02,Dloss03,Coish2004,Dassarma2009,Coish2010}, the Bethe ansatz method for the polarized bath with one flipped  bath spin \cite{Bortz07}, hybrid methods combining the Bethe ansatz and Monte Carlo sampling \cite{AFprl,Alexandre2013}, the DMRG method and semiclassical approaches \cite{Stanek2013,Stanek2014,Schering2019,Schering2020}.

Recent work  described in detail the collapse and revival dynamics of the homogeneous central spin model \cite{prbxxz}, and the partition of eigenstates between entangled and separable states \cite{claey20,ningwu}.

 While an early description of Gaudin magnets using the Bethe ansatz technique considered one 
flipped bath spin, i.e. $M=1$ \cite{Bortz07}, here we study the time evolution of the central spin problem  with an  arbitrary number of flipped spins $M$, starting from an 
 initial state $\vert \Phi_{0} \rangle=s_{a_{1}}^{-} \cdots s_{a_{M}}^{-} \vert \Uparrow \rangle=\vert a_{1},\cdots,a_{M}\rangle$, where $\vert \Uparrow \rangle$ is the fully polarized state and $a_i = 0,1,2, ... N$ label the positions of spins in the system.    
The quantum dynamics of the model  with  arbitrary number of flipped spins has been already studied through Bethe ansatz \cite{AFprl,Alexandre2013}, as well as various other theoretical approaches, leading to important insights such as a description of quantum decoherence by the Chebyshev expansion method \cite{Dobrovitski}, non-Markovian dynamics \cite{Coish2004}, persistent spin correlations \cite{Uhrig2014}, etc. 
Here we study the time evolutions of several interesting quantities analytically and numerically: the spin distribution function, the spin-spin correlation function, and the Loschmidt echo. Evaluating these, we find that the time dependence of the spin polarization function reveals a breakdown of thermalization, such that the steady-state of the Gaudin magnets cannot be described by an equilibrium ensemble. 
Instead, the relaxation to steady-state values of the spin-spin correlation function suggests a logarithmic decay, and we find that at intermediate times the Loschmidt echo relaxes exponentially to its steady-state value, using both the Bethe ansatz solution and Matrix Product States (MPS) numerical approach.


\section{Model and Bethe ansatz equations}
The Gaudin magnet describes a central spin  at position ``$0$'' coupled to bath spins through long-range interactions. The Hamiltonian reads
\begin{equation}
H=B\mathbf{s} ^{z}_{0}+2 \sum_{j=1}^{N}A_{j}\textbf{s}_{0} \cdot\textbf{s}_{j},
\label{Hami}
\end{equation}
where the bath spins are labelled with $1 \rightarrow N$. For convenience, we parametrize the magnetic field as $B=-{2}/{g}$. Furthermore, we write $ A_{j}=1/(\epsilon_{0}-\epsilon_{j})$  for the inhomogeneous interaction strength determined in experiments. Setting $ \epsilon_{0}=0$, the  $\epsilon_{j}$ (with $j=1,\ldots, N$) correspond to the energy levels of a discrete BCS model associated with $H$, which can be constructed from the conserved quantities $H_{j}=B \textbf{s}_{j}^{z}+\sum_{k \neq j}\textbf{s}_{j} \cdot\textbf{s}_k/(\epsilon_{j}-\epsilon_{k})$ \cite{J.Dukelsky2004,H.-Q. Zhou,Guan-2002}.  
Following the  algebraic Bethe ansatz method, and considering a subspace where $M$ spins are flipped-down with respect to the reference state $\vert \Uparrow \rangle$, the eigenstates are given below \cite{J.Dukelsky2004,H.-Q. Zhou}:
\begin{equation}
\vert \nu_{1},\cdots,\nu_{M} \rangle=\prod_{\alpha=1}^{M}B_{\nu_{\alpha}} \vert \Uparrow \rangle=\prod_{\alpha=1}^{M}\sum_{j=0}^{N} \frac{s_{j}^{-}}{\nu_{\alpha}-\epsilon_{j}}  \vert \Uparrow \rangle,
\label{EigF}
\end{equation}
where the $M$ parameters $\lbrace \nu_\alpha\rbrace$ should satisfy the following Bethe ansatz equations
\begin{equation}
 \sum \limits_{j=0}^{N} \frac{1}{\nu_{\alpha}-\epsilon_{j}}=\frac{2}{g}+\sum_{\beta \neq \alpha , \beta=1}^{M} \frac{2}{\nu_{\alpha}-\nu_{\beta}},
 \label{BAE}
\end{equation}
with $ \alpha =1, 2, \ldots, M$. The equations (\ref{BAE}) are also  called Richardson-Gaudin equations. There are $C_{N+1}^{M}$ sets of solutions to Eq.~(\ref{BAE}), which correspond  to the number of choices of flipping $M$ spins \cite{AFprb}. These $C_{N+1}^{M}$ states $ \vert \nu_{1},\cdots,\nu_{M} \rangle$ span the subspace with $M$ spins flipped-down.
In Appendix D we give details of the numerical method for solving the Bethe Ansatz Eq.~(\ref{BAE}). Moreover, their eigenenergy is given by 
\begin{equation}
  E=\frac{B}{2}+\frac{1}{2} \sum_{j=1}^{N} \frac{1}{\epsilon_{0}-\epsilon_{j}}-\sum_{\alpha=1}^{M}\frac{1}{\epsilon_{0}-\nu_{\alpha}}.
  \label{Energy}
 \end{equation}
 
We exploited the above eigenfunctions to study the quantum dynamics of the central spin model. The initial state (see introduction) is a simple product state $\vert \Phi_{0} \rangle=\vert a_{1},\cdots,a_{M}\rangle$, where $\{ a_{1},\cdots,a_{M} \}$  is a set of mutually unequal site indexes which can range from ``0'' to ``$N$''. After the interaction is switched on, the wave function evolves as
\begin{equation}\label{psi_t}
\vert \psi(t) \rangle= \sum_{k} \vert \phi_{k} \rangle \langle \phi_{k} \vert \Phi_{0} \rangle  e^{-iE_{k}t},
\end{equation}
where we introduced the orthonormalized wave functions $\vert \phi_{k} \rangle=N_{\nu_{k}} \vert \nu_{1,k},\cdots,\nu_{M,k} \rangle$, with $N_{\nu_{k}}$ the normalization coefficient\cite{SuppM} (see Appendix~\ref{appendix_sz}). The index ``$k$'' labels all the states corresponding to the roots of Eq.~(\ref{BAE}). Since the initial state has $M$ spins which are flipped-down, we only need to consider the Bethe ansatz basis in  this particular subspace. A critical quantity in Eq.~(\ref{psi_t}) is the overlap between the initial state and the eigenfunctions, given as follows:
\begin{equation}\label{Phi_0}
\langle \Phi_{0} \vert \phi_{k} \rangle=N_{\nu_{k}} \sum_{\mathcal{P}\in \lbrace a_{1},\cdots,a_{M} \rbrace} \prod \limits_{\alpha=1}^{M} \frac{1}{(\nu_{\alpha,k}-\epsilon_{\mathcal{P}_{\alpha}})},
\end{equation}
where ``$\mathcal{P}$'' means summing over all permutation of indexes  $\lbrace a_{1},\cdots,a_{M} \rbrace$, see Appendices A and B. The overlap can be also written as $ \langle \Phi_{0} \vert \phi_{k} \rangle=N_{\nu_{k}} \perm(1/(\nu_{\alpha,k}-\epsilon_{\mathcal{P}_{\alpha}})$, where ``$\perm$" is the permanent of matrix $1/(\nu_{\alpha,k}-\epsilon_{\mathcal{P}_{\alpha}})$, similar to the determinant \cite{J.Dukelsky2004,H.-Q. Zhou}. It is worth noting that both permanent and determinant representation can give the exact quantum dynamics of the Gaudin magnets. However, the numerical tasks from the two representations is quite different as determinants can be evaluated in polynomial time. The interested readers can find more details about permanent  and determinant representations  in \cite{caux2008,AFJPA,Claeys2017}. For either representation, it still remains challenging to get 
dynamical quantities for the Gaudin magnets with a large system size, and values up to $N = 48$ were reached \cite{AFprl}.

From expressions (\ref{psi_t}) and (\ref{Phi_0}), the evolution of observables (for example the spin distribution) can be calculated rather straightforwardly. Based on this approach, we will present in the rest of the paper our key analytical and numerical results on the quantum dynamics of Gaudin magnets. 

\section{Spin distribution}

By making use of Eq.~(\ref{psi_t}), the time-dependent polarization of site $j$ is immediately written as
 \begin{equation}
s^{z}_{j}(t)=\sum_{k} \sum_{k^{\prime}} \langle \Phi_{0} \vert \phi_{k} \rangle  \langle  \phi_{k}  \vert s_{j}^{z} \vert \phi_{k^{\prime}}  \rangle \langle  \phi_{k^{\prime}}  \vert \Phi_{0} \rangle e^{-i w_{kk^{\prime}}t},
\label{Szt}
\end{equation}
where $w_{kk^{\prime}}=E_{k^{\prime}}-E_{k}$ simplifies to $\sum_{\alpha=1}^{M}1/(\epsilon_{0}-\nu_{\alpha,k})-\sum_{\alpha=1}^{M} 1/(\epsilon_{0}-\nu_{\alpha,k^{\prime}})$. While the overlaps are given in Eq.~(\ref{Phi_0}), the remaining difficulty is to obtain the matrix elements of the observable in the eigenfunction basis. As detailed in Appendix B, one can insert the complete set of states $\vert a_{1},\cdots,a_{M}\rangle$, thus writing the matrix elements in terms of overlaps of type $\langle \Phi_{0} \vert \phi_{k} \rangle$. We finally obtain the exact expression:
\begin{align}
s_{j}^{z}(t)=&\frac{1}{2}-\sum_{k}\sum_{k^{\prime}} \cos(w_{kk^{\prime}}t) \sum\limits_{ j_{1}< \cdots <j_{M}} \sum_{\alpha=1}^{M}\delta_{jj_{\alpha}}\nonumber \\
&\times  \sum_{\mathcal{P}}\sum_{\mathcal{Q}} \frac{\vert N_{\nu_{k}}\vert^{2}}{\prod \limits_{\alpha=1}^{M}(\nu_{\alpha,k}-\epsilon_{{\mathcal{P}_{\alpha}}}) (\nu_{\alpha,k}-\epsilon_{{\mathcal{Q}_{\alpha}}})} \nonumber \\
&\times  \sum_{\mathcal{P}}\sum_{\mathcal{Q}} \frac{\vert N_{\nu_{k'}}\vert^{2}}{\prod \limits_{\alpha=1}^{M}(\nu_{\alpha,k'}-\epsilon_{{\mathcal{P}_{\alpha}}})(\nu_{\alpha,k'}-\epsilon_{{\mathcal{Q}_{\alpha}}})} ,
\label{sjz}
\end{align}
where, here and throughout the article, ``$\mathcal{P}$" and ``$\mathcal{Q}$" indicate summing over all the permutations of indexes $ \lbrace a_{1},\cdots,a_{M} \rbrace$ and $ \lbrace j_{1},\cdots,j_{M} \rbrace$, respectively. 
	 
Although it is not immediately obvious from Eq.~(\ref{sjz}), $s_{j}^{z}(t)$ recovers the correct initial value when we take the limit  $t \rightarrow 0$. Detailed proof can be found in Appendix~\ref{appendix_sz}. There we also show that, as expected, $m^{z}=\sum_{j}s_{j}^{z}(t)=(N+1)/2-M$ is conserved by Eq.~(\ref{sjz}). The evolution of spin distribution function for a more general initial state is just  an extension  of above result Eq.~(\ref{sjz}). 

\begin{figure}
\begin{center}
\includegraphics[width=0.45\textwidth]{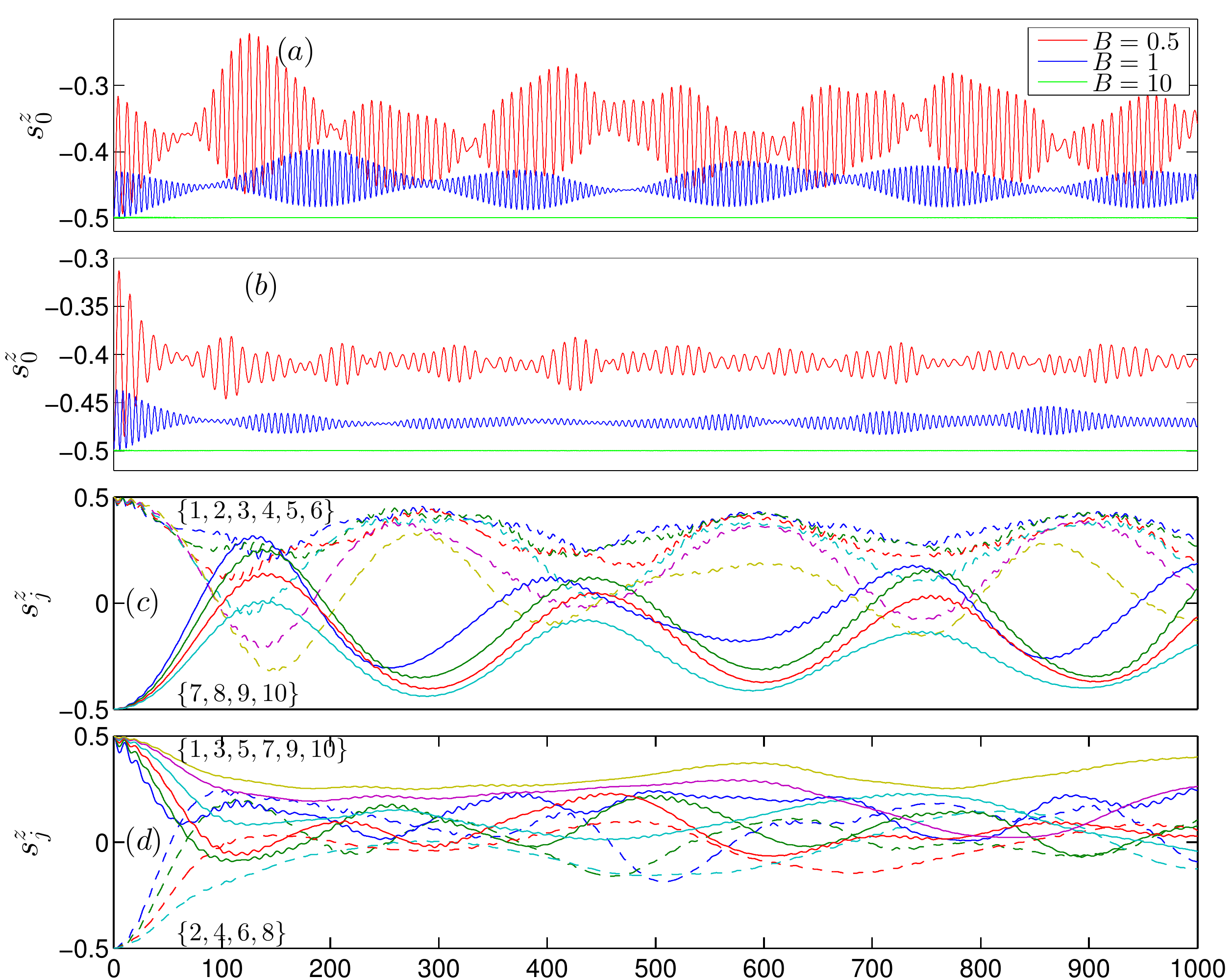}
\end{center} 
\caption{Time evolution of the spin polarization of individual spins, obtained from Eq.~(\ref{BAE}) assuming an exponentially decaying coupling constant $A_{j}=A/N \exp(-j/N)$. We take $N=10$ and use dimensionless units, setting $A=1$. In (a)(c) we choose an initial state $\vert \Phi_{A} \rangle=\vert 0,7,8,9,10\rangle$. In (b)(d) the initial state is $\vert \Phi_{B} \rangle=\vert 0,2,4,6,8\rangle$. Figures~(a) and (b) present the dynamical polarization of central spins with different magnetic fields for the two initial states.
Figures~(c) and (d) show the dynamical evolution of  bath spins at $B=0.5$ for the two initial states. 
}
\label{fig1sz0}
\end{figure}

\begin{figure}
  \begin{center}
\includegraphics[width=0.45\textwidth]{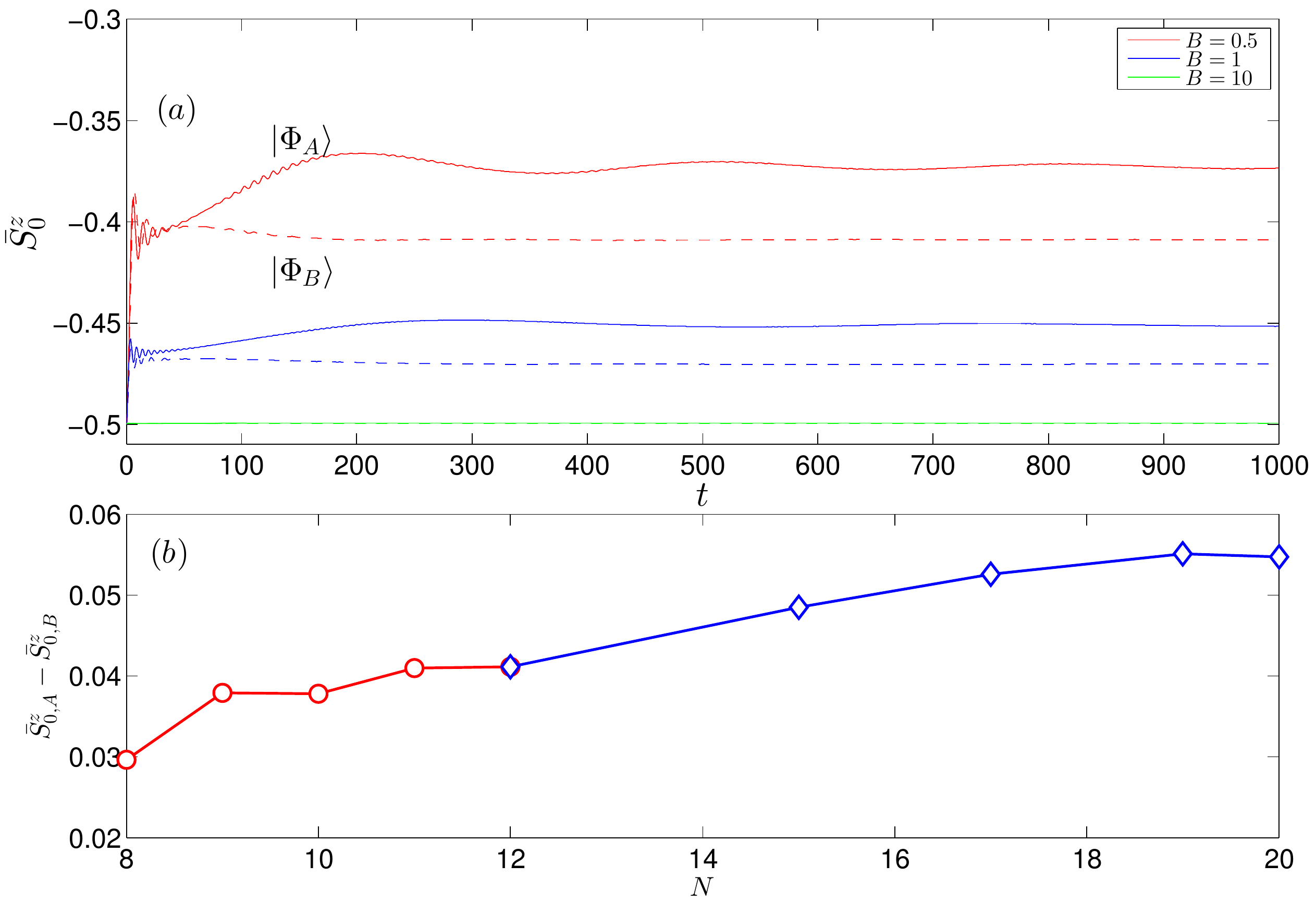}
\end{center} 
\caption{(a) Time average of the central spin polarization, computed for different values of $B$ and $N=10$. The solid lines are for the  initial state $\vert \Phi_{A} \rangle$ while the dashed lines are for the initial state $\vert \Phi_{B} \rangle$. (b) Dependence of $\bar{s}_{0,A}^{z}-\bar{s}_{0,B}^{z}$ (i.e., the difference in the time-averaged central spin polarization between the two initial conditions) on the bath spin number $N$. Here we use a fixed value of  $B=0.5$. The red circles are from the exact Bethe ansatz equations, whereas the  blue diamonds are obtained by the Matrix Product State (MPS) method (see Sec.~\ref{sec:Loschmidt}).}
\label{figtsz0}
\end{figure}

\begin{figure*}
\includegraphics[scale=0.4]{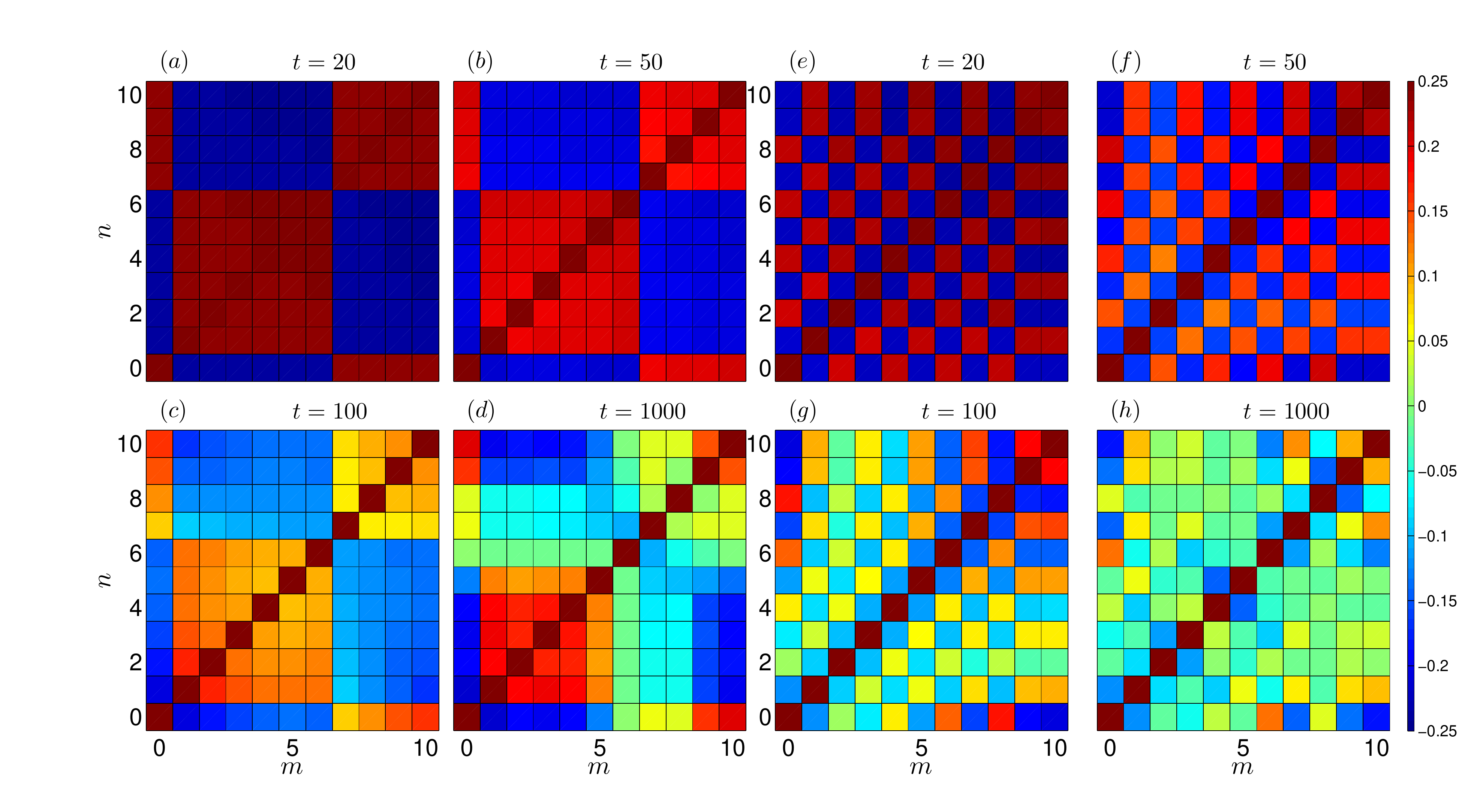}
\caption{Left panels: the spin-spin correlation function $ G_{mm}^z$ at different times, with a bath size $N=10$, magnetic field $B=0.5$, and initial state $\vert \Phi_{A} \rangle=\vert 0,7,8,9,10 \rangle$. Right panels:
spin-spin correlation function $ G_{mm}^z$ with the initial state $\vert \Phi_{B} \rangle=\vert 0,2,4,6,8 \rangle$.}
\label{figszsz}
\end{figure*}

We have evaluated Eq.~(\ref{sjz}) for two representative initial conditions, i.e., the domain-wall state $\vert \Phi_{A} \rangle = s_{0}^{-} \left(\prod_{a> N/2+1} s_{a}^{-} \right)\vert \Uparrow \rangle $ and the anti-ferromagnetically ordered state $\vert \Phi_{B} \rangle = \prod_{a=0}^{N/2-1} s_{2a}^{-} \vert \Uparrow \rangle$, and present in Fig.~\ref{fig1sz0} the resulting polarization of individual spins. The dynamical evolution of the central spin is shown in panels (a) and (b), and  is allowed by the interaction with the bath spins, which leads to change of the orientation of the central spin through flip-flop processes. It is characterized by a  quick decay to the steady-state value, with the persistence of a oscillatory behavior due to finite-size effects \cite{AFprl}. There is a competition between magnetic field and coupling to the bath spins, evident from the evolution curves at different values of $B$ shown in Fig.~\ref{fig1sz0}(a)  and (b). The physical mechanism behind the suppressed evolution at large $B$ is the increase of the energy gap between states with ${\bf s}^z_0=\pm 1/2$, which inhibits changes of orientation of the central spin. In Appendix~\ref{sec:VNE}, we also present the detailed derivation of the reduced density matrix and von Neumann entropy for the central spin. The general behavior of the central spin polarization in Fig.~\ref{fig1sz0} is very similar to the oscillatory time dependence observed for the von Neumann entropy, see Fig.~\ref{fig:VNE}. 

For models with short-range interactions, it is well-known that the time-evolution of non-equilibrium states is characterized by a light-cone dynamics of local observables~\cite{natan2014,lieb72,Ibloch,caux2014}. On the other hand, the Gaudin magnets is a long-range model and, as shown in Fig.~\ref{fig1sz0}(c) and (d),  the spin polarization of the bath spins evolve simultaneously into their steady-state value. Being infinite-range, the interaction of the Gaudin magnet can take effect without any propagation time. 

Comparing the two initial states,  we see that with $\vert \Phi_{A} \rangle$ the bath spins need a much longer time than $\vert \Phi_{B} \rangle$  to evolve into a steady state. In fact, for the initial state $\vert \Phi_{A}\rangle$ the two groups of bath spins remain well distinct, and the oscillation in the spin polarization nearly approach periodically their initial values ($1/2$ or $-1/2$). On the other hand, the polarizations for the initial state $\vert \Phi_{B}\rangle$ show an initial quick decay, leading to oscillations around  $0$. The evolution for the initial  state $\vert \Phi_{A} \rangle$ displays a strong memory of the initial domain-wall order.  In order to further understand such peculiar dynamics,  in later sections we will study the time evolution of the spin-spin correlations and the Loschmidt echo. 

If the time-average $\bar{\mathcal{O}}$ of an observable coincides its micro-canonical ensemble average, i.e., $\bar{\mathcal{O}}=\langle \mathcal{O} \rangle$, the system reaches thermalization. Here the time average of the spin distribution $\bar{s}_{j}^{z}(t)=1/t\int_{0}^{t}s_{j}^{z}(\tau)d\tau$ can be simply obtained from Eq.~(\ref{sjz}), by replacing $ \cos(w_{kk^{\prime}}t)$ with $ \sinc(w_{kk^{\prime}}t)=\sin(w_{kk^{\prime}}t)/w_{kk^{\prime}}t$. We see in Fig.~\ref{figtsz0}(a) that the time-average of spin polarization of the central spin converges to clearly different values for the two initial states. The difference in the steady-state spin polarizations does not  decrease by increasing $N$, see Fig.~\ref{figtsz0}(b), in which smaller values of $N$(red circles) are obtained by exact method, relatively larger values of $N$(blue diamond) are obtained by MPS method. The combined results of BA and MPS method verify the persistence of a finite difference $\bar{s}_{0,A}^{z}-\bar{s}_{0,B}^{z}$ for the steady-state value. This observation confirms a  breakdown of thermalization in the  inhomogeneous central spin model, due to the restricted time evolution implied by integrability. 
Since the existence of conserved quantities, like $H_{j}=B \textbf{s}_{j}^{z}+\sum_{k \neq j}\textbf{s}_{j} \cdot\textbf{s}_k/(\epsilon_{j}-\epsilon_{k})$, the time-evolution cannot be ergodic, neither in the whole Hilbert space nor in a subspace with fixed $M$.

\section{Spin-Spin correlations}

The spin-spin correlation function is defined as $G^{z}_{mn}(t)=\langle \psi(t) \vert s_{m}^{z}s_{n}^{z}\vert \psi(t) \rangle$ and, as discussed in Appendix~\ref{appendix_sz}, which can be computed in a way similar to the spin density. The explicit expression reads:
\begin{align}
G^{z}_{mn}(t)& =\sum_{k,k'} \sum\limits_{ j_{1}< \cdots <j_{M}} \left(\frac{1}{2}-\sum \limits_{\alpha=1}^{M}\delta_{mj_{\alpha}}\right)\left(\frac{1}{2}-\sum \limits_{\alpha=1}^{M}\delta_{nj_{\alpha}}\right) \nonumber \\
&\times   \sum_{\mathcal{P},\mathcal{Q}}\frac{\vert N_{\nu_{k}}\vert^{2}}{\prod \limits_{\alpha=1}^{M}(\nu_{\alpha,k}-\epsilon_{{\mathcal{P}_{\alpha}}}) (\nu_{\alpha,k}-\epsilon_{{\mathcal{Q}_{\alpha}}})} \nonumber \\
&\times  \sum_{\mathcal{P},\mathcal{Q}} \frac{\vert N_{\nu_{k'}}\vert^{2}}{\prod \limits_{\alpha=1}^{M}(\nu_{\alpha,k'}-\epsilon_{{\mathcal{P}_{\alpha}}}) (\nu_{\alpha,k'}-\epsilon_{{\mathcal{Q}_{\alpha}}})}\cos(w_{kk^{\prime}}t)  ,
 \label{Eszsz}
\end{align}
 which characterizes longitudinal correlation between spins. 

In Fig.~\ref{figszsz}, we show the spin-spin correlation function at different times, both for the initial state $\vert \Phi_{A} \rangle$ (left panels) and $\vert \Phi_{B} \rangle$ (right panels). On a relatively short timescale ($t<50$), the correlation function is very similar to $t=0$ for both initial states and  when the evolution time approaches  $t\sim 10^{2}$ the correlation function has decreased to smaller values. However, it can be noted that in the left panels of Fig.~\ref{figszsz} (referring to $\vert \Phi_{A} \rangle$), the domain-wall order has not faded away completely even after a long evolution time $t\sim 10^3$. This is in contrast to panel (h) on the right side, where the anti-ferromagnetic alignment has nearly disappeared. This confirms our previous observation that the domain-wall configuration is more favourable to retain memory of the initial state (see also Fig.~\ref{lechoBC} for the Loschmidt echo).

From the contour plot of the correlations in Fig.~\ref{figszsz},  the correlations between the  bath spins $G_{mn}^z$ with $m,n>0$ tend to zero. Meanwhile the correlations between the central spin and bath spins  $G_{m0}^z$ with $m>0$  oscillate near initial value. After long enough time evolution ($t\sim 10^{3}$), the correlations between the  bath spins almost evolve toward to zero for  the state $\vert \Phi_{B} \rangle$.
 
Considering in more detail the right side of Fig.~\ref{figszsz}, we see in panel (g) that the correlations $G_{mn}^z$ with $m,n \gtrsim 5$ are more robust, and also the values of $G_{0n}^z$ with larger $n$ remain rather similar to the initial values. This behavior is a consequence of the small coupling strength $A_j$ when $j$ is large, which implies a longer timescale for those bath spins. However, after a sufficiently long time evolution ($t\sim 10^{3}$) the correlations between all spins are close to zero. The vanishing of correlations in panel (h) indicates that the steady state is like a paramagnetic state.

\begin{figure}
  \begin{center}
\includegraphics[width=0.45\textwidth]{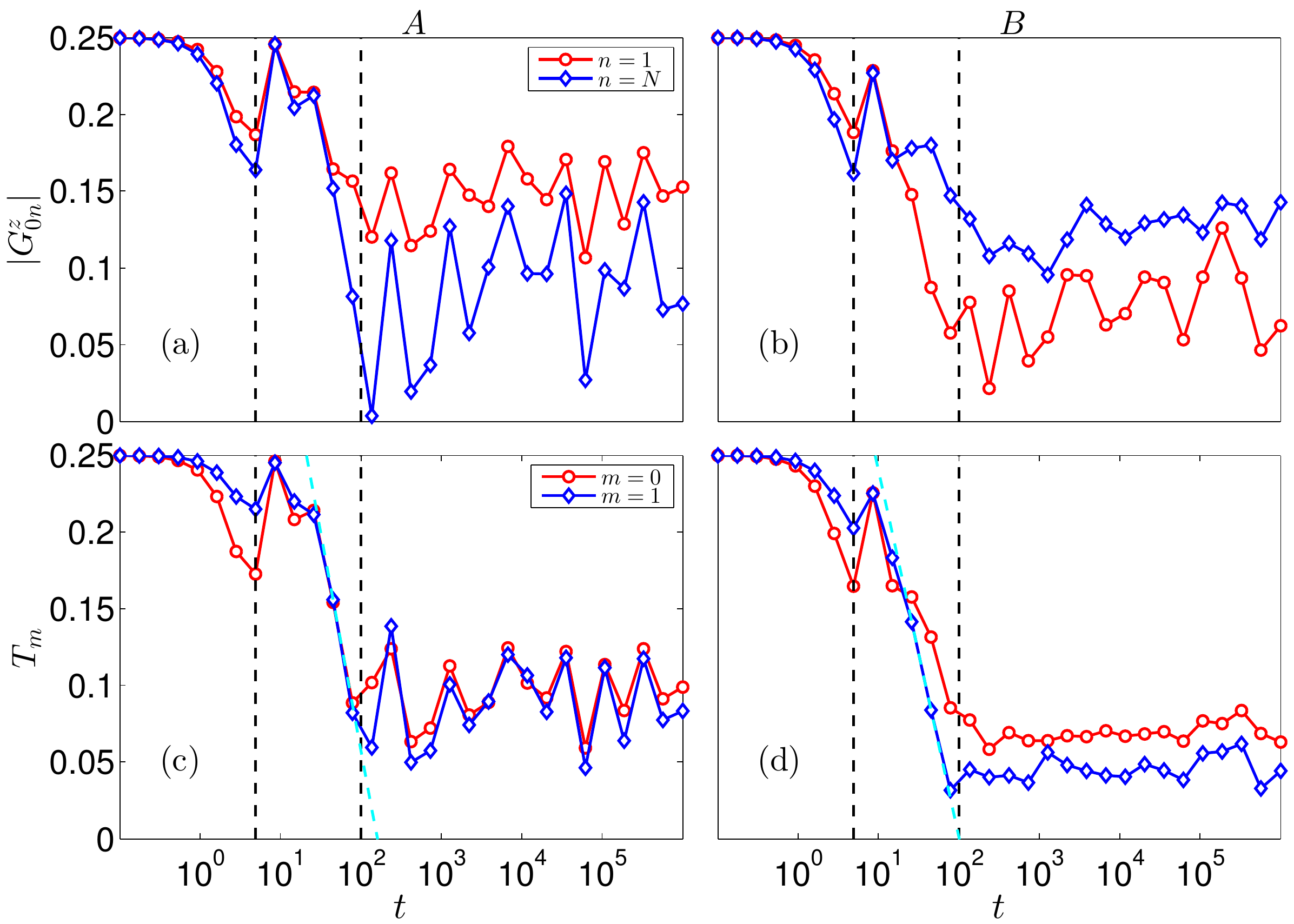}
\end{center} 
\caption{Panels (a) and (b) plot the spin-spin correlation function $\vert G_{0n}^{z}\vert$ for spins $n=1,N$. Panles (c) and (d) average the spin-spin correlation function according to Eq.~(\ref{meanszz}). The bath size is $N=10$ and the magnetic field $B=0.5$. Left panels: initial state $\vert \Phi_{A} \rangle=\vert 0,7,8,9,10 \rangle$. Right panels: initial state $\vert \Phi_{B} \rangle=\vert 0,2,4,6,8 \rangle$.}
\label{figmeanszz}
\end{figure} 

To further analyze the evolution of $G_{mn}^z(t)$, we characterize the typical correlation strength for spin $m$ as
 \begin{equation}
 T_{m}=\sqrt{\frac{\sum \limits_{n\neq m}(G_{mn}^z(t))^{2}}{N}},
 \label{meanszz}
 \end{equation}
where the self-correlation terms $G_{mm}^z$ are excluded. In Fig.~\ref{figmeanszz}, we show the absolute value of correlation function and the typical correlation strength for the two initial states, i.e., the left two panels (a,c) are for the initial state $\vert \Phi_{A} \rangle$ and the right two panels (b,d) for $\vert \Phi_{B} \rangle$. For both ferromagnetic and anti-ferromagnetic alignment, the correlation strength decays from initial maximum value of $1/4$ towards smaller values, with $T_m =0$ corresponding to disordered spin alignment. We can see that there are two stages in  the relaxation of correlations. After the initial decay (up to $t \lesssim 10$, see the first black dash line), there is an intermediate regime in which the correlation strength has a logarithmic relaxation (up to $t\sim 10^{2}$, near the  second black dash line), i.e., $\Delta T_{m}\propto -\ln t$. In this interval the linear decay shown in Fig.~\ref{figmeanszz} (on a semi-logarithmic scale) has its steepest slope. Finally, in the long-time limit the central spin is strongly entangled with bath spins and the value of $T_m$ reaches its steady-state value. Due to finite-size effects, the decay is modified by oscillations, see especially panel (b), but the qualitative form and timescales are quite robust with respect to the value of $m$ and the type of initial state.

 \section{Loschmidt echo and MPS approach}\label{sec:Loschmidt}
The Loschmidt echo, defined as $L(t)=\vert \langle \Phi_{0} \vert \psi(t) \rangle \vert^2$, quantifies the memory of the initial state \cite{heyl2013}, thus can provide us a clear  understanding of  the difference between the evolution for the two initial states. By using Eq.~(\ref{psi_t}), we express $L(t)$ as:
\begin{eqnarray}
L(t)=\sum_{k} \sum_{k^{\prime}}\vert \langle \Phi_{0} \vert \phi_{k} \rangle \vert^2  \vert \langle  \phi_{k^{\prime}}  \vert \Phi_{0} \rangle \vert^2 e^{-i w_{kk^{\prime}}t},
\end{eqnarray}
where the overlaps are given in Eq.~(\ref{Phi_0}). Finally, we obtain:
\begin{align}
L(t)=&\sum_{k}\sum_{k^{\prime}}\left[ \sum_{\mathcal{P}}\frac{\vert N_{\nu_{k}}\vert}{\prod \limits_{\alpha=1}^{M}(\nu_{\alpha,k}-\epsilon_{\mathcal{P}_{\alpha}})} \right]^{2}  \nonumber \\
&\times \left[ \sum_{\mathcal{P}}\frac{\vert N_{\nu_{k^{\prime}}}\vert}{\prod \limits_{\alpha=1}^{M}(\nu_{\alpha,k^{\prime}}-\epsilon_{\mathcal{P}_{\alpha}})} \right]^{2} \cos(w_{kk^{\prime}}t).
\label{Lecho}
\end{align}

\begin{figure}
\begin{center}
\includegraphics[scale=0.4]{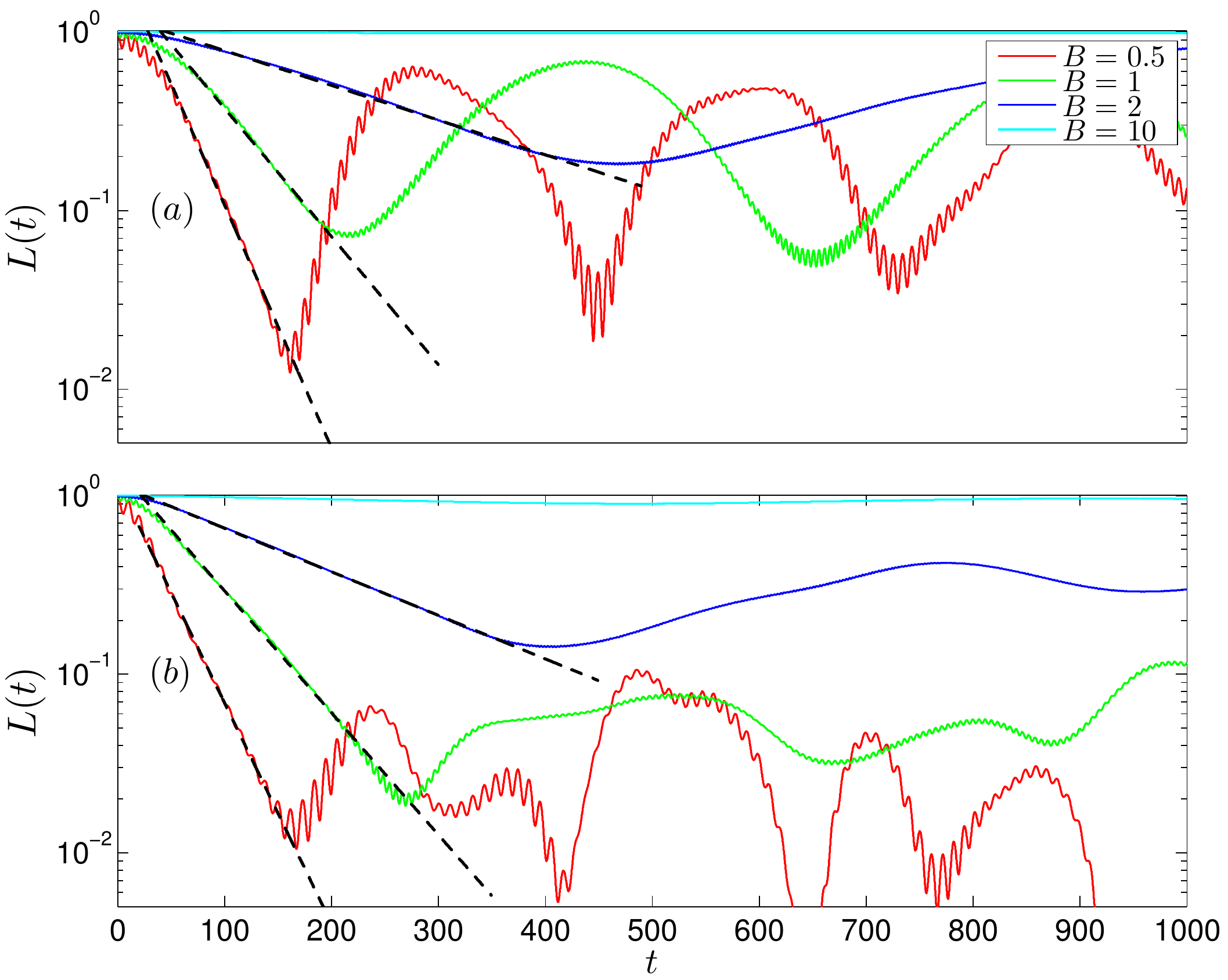}
\end{center} 
\caption{Time dependence of the Loschmidt echo for two different initial state. Panel (a) with the initial state $\vert \Phi_{A} \rangle$, and panel (b) with initial state $\vert \Phi_{B} \rangle$.}
\label{lechoBC}
\end{figure} 

We show representative examples of the Loschmidt echo in Fig. \ref{lechoBC}. As seen in panel (a), for a  weak magnetic field ($B\sim 0.5$) the initial state $|\Phi_{A}\rangle$ leads to a long-time evolution with large oscillations around a mean value $\bar{L}(t)\sim 0.25$. Instead, panel (b) shows that $L(t) \sim 0$ in the long-time limit, thus the system nearly loses its memory of the initial state  $|\Phi_{B}\rangle$. 

In the first part of the time evolution, besides the presence of fast small amplitude oscillations, the Loschmidt echo displays a clear exponential dependence $\Delta L(t) \propto \exp(-\gamma t)$, up to an intermediate time scale $t\sim10^{2}$. According to our numerical calculation, the decay coefficients is similar for both initial states, i.e., $\gamma_{A}\simeq \gamma_B \sim 0.03 $ with $B=0.5$. Such exponential scaling behavior of the Loschmidt echo has been widely found in dynamical evolution of quantum many-body system \cite{Carla,fazio,Dubertrand,Zurek20}. 

\begin{figure}
\begin{center}
\includegraphics[width=0.45\textwidth]{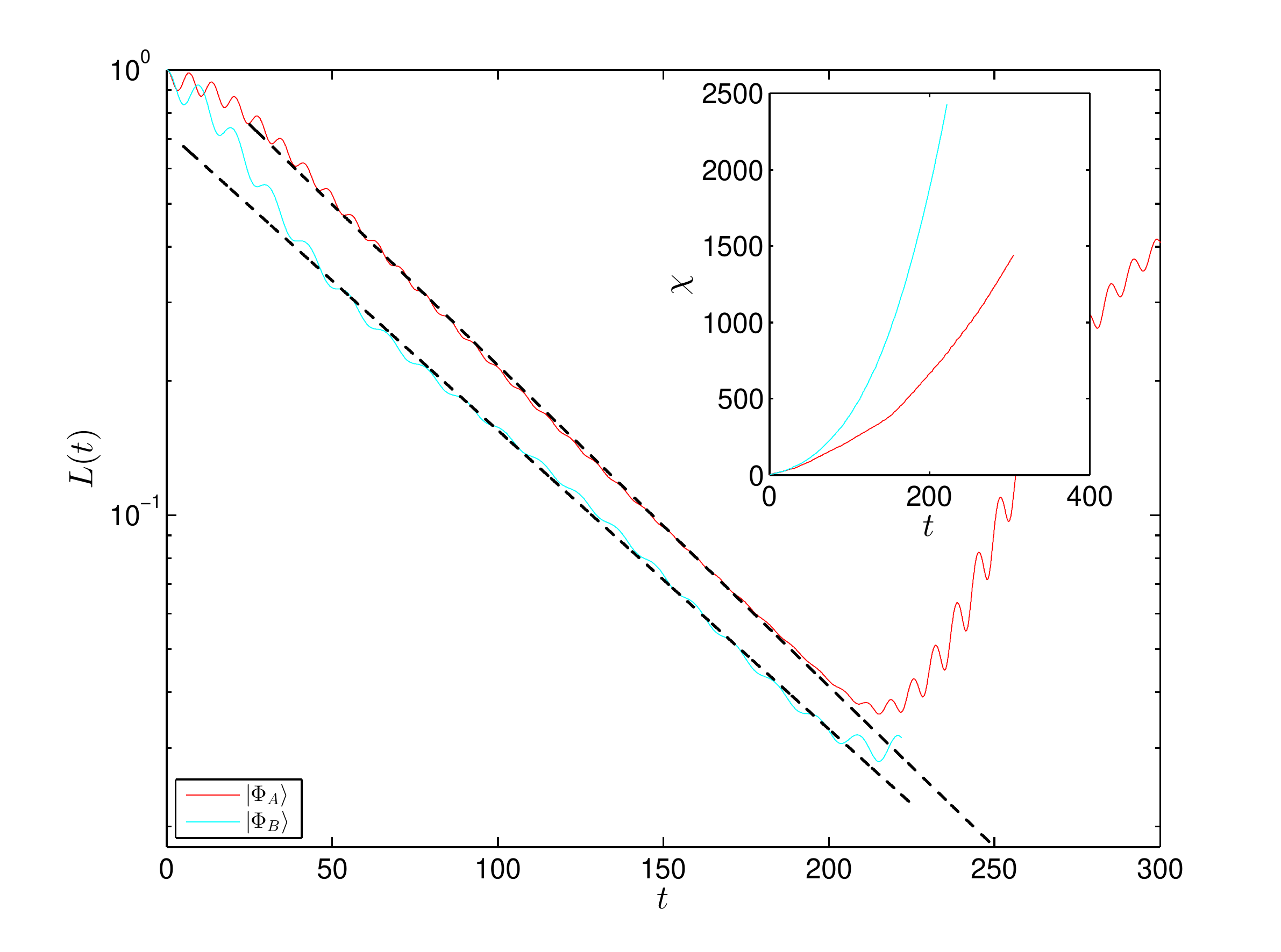}
\end{center} 
\caption{ Loschmidt echo as function of time $t$ for two different initial states by using the MPS method with $N=50$. The red line is for the initial states $\vert \Phi_{A} \rangle$, and the cyan line is for the initial state $\vert \Phi_{B} \rangle$.}
\label{lechomps}
\end{figure}

In order to confirm  the scaling relaxation behavior of Loschmidt echo for a larger system size, we further perform an approximate calculation based on the MPS method \cite{Gvidal,Schollwock}, as implemented in the Itensor library \cite{itensor}. There are two important input parameters of the MPS simulations: the variational tolerance $\epsilon$ of the MPS wavefunction, which represents the error of the MPS state in approximating the true wavefunction  (we use  $\epsilon=10^{-9}$ in our simulations), and the bond dimension $\chi$, which indicates the maximum dimension of the matrices entering the variational ansatz. In the given implementation of the MPS method, the bond dimension is automatically increased to meet the desired tolerance. In Fig.~\ref{lechomps}, we show the Loschmidt echo for a bath size $N=50$ and the two types of initial states considered in this work, $\vert \Phi_{A} \rangle$ (red lines) and $\vert \Phi_{B} \rangle$ (cyan lines). Also in this case the Loschmidt echo is characterized by an initial exponential decay, and this behavior persists up to intermediate times ($t \approx 10^{2}$). 

Unfortunately, we are not able to access the long-time dynamics at $N=50$ based on the MPS approach, as the entanglement in the wavefunction grows with time. This reflects itself on a rapid increase of the bond dimension $\chi$, which eventually reaches numerically intractable values. The time dependence of $\chi$ in the simulation of Fig.~\ref{lechomps} is shown in the inset. As seen, the bond dimension $\chi$ of initial state $\vert \Phi_{B} \rangle$ grows faster in time than $\vert \Phi_{A} \rangle$. Nevertheless, the range of times we can simulate is sufficient to establish conclusively the initial exponential form of decay of Loschmidt echo.
In classical systems, the exponential decay of Loschmidt echo usually implies chaotic dynamics \cite{Dubertrand,Prosen,Weinstein}.  For the central spin model, the exponential decay of Loschmidt echo may indicate the dynamics is chaotic here as well \cite{Emerson}.

\section{Conclusion}
In summary, based on Bethe Anstaz techniques, we have analytically studied the time evolution of several dynamic quantities of Gaudin magnets with arbitrary number $M$ of flipped-down spins, such as spin distribution function, spin-spin correlation function, and Loschmidt echo. Furthermore, we have numerically studied the scaling behavior of the relaxation dynamics, up to relatively large system sizes. Significantly, the correlation function relaxes to its steady value with a logarithmic dependence, and the Loschmidt echo reveals an exponential  loss of the memory of the initial state. Our results highlight several interesting features of the dynamical evolution of a quantum many-body system with long range interactions, thus advancing the understanding of decoherence for single qubits coupled to a spin bath at the many-body level. With respect to universal features of quantum many-body systems, much effort is still necessary to better understand the non-equilibrium quantum dynamics \cite{suter}. In future studies, it would be desirable to extend our methods to investigate scrambling in many-body system~\cite{htshen,prxotoc}, and the out-of-time-of-correlation function (OTOC)~\cite{mblotoc} of central-spin models. The details of such dynamical evolution of spin correlations may be accessible experimentally, e.g., through NMR techniques \cite{niknam}.

\acknowledgements

We would like to thank R. Fazio for useful discussions and F. Iemini and D. Ferreira their kind help with the MPS simulation package. W.B.H. acknowledges support from NSAF (Grant No. U1930402). X.W.G. is supported by the key NSFC grant No.\ 11534014 and No.\ 11874393, and the National Key R\&D Program of China  No.\ 2017YFA0304500. S.C. acknowledges support from NSFC (Grants No. 11974040 and No. 1171101295) and the National Key R\&D Program of China No. 2016YFA0301200. H. Q. L. acknowledges financial support from National Science Association Funds U1930402 and NSFC 11734002, as well as computational resources from the Beijing Computational Science
Research Center.


\begin{appendix}

\section{Bethe ansatz basis}

The eigenstates of Eq.~(\ref{Hami}) can be obtained with Bethe ansatz (BA) approach, as discussed in detail in Ref.~\onlinecite{H.-Q. Zhou}. Here we discuss several properties of these states which will find wide use in the following derivations. Since the Hilbert space of the Gaudin model can be decomposed into the direct sum of subspaces  with $M$  flipped-down spins, we introduce two basis sets for such subspaces. One is the natural basis $\vert   j_{1},\cdots,j_{M}  \rangle=S_{j_{1}}^{-},\cdots,S_{j_{M}}^{-} \vert \Uparrow \rangle$, and another is the Bethe ansatz basis $\vert \nu_{1,k},\cdots,\nu_{M,k} \rangle=B_{\nu_{M,k}},\cdots,B_{\nu_{1,k}}  \vert \Uparrow \rangle$. The corresponding completeness relations are simply given by:
\begin{align}
\sum_{M} \sum_{ j_{1},\cdots,j_{M}}  \vert   j_{1},\cdots,j_{M}  \rangle\langle j_{1},\cdots,j_{M} \vert =I, \\
\sum_{M,k}   \vert N_{\nu_{k}}\vert^{2} \vert \nu_{1,k},\cdots,\nu_{M,k} \rangle  \langle  \nu_{1,k},\cdots,\nu_{M,k}\vert  =I,
\end{align}
where $\vert N_{\nu_{k}}\vert$ are the normalization coefficients for the Bethe ansatz basis. The two sets of states $\vert   j_{1},\cdots,j_{M}  \rangle$ and $\vert \nu_{1,k},\cdots,\nu_{M,k} \rangle$ are both complete in the subspace with $M$ flipped-down spins. Including all the $M$, the complete basis of the whole Hilbert space is obtained.

The explicit form of the Bethe ansatz eigenfunctions reads 
\begin{align}
\vert \nu_{1,k},\cdots,\nu_{M,k} \rangle =\prod_{\alpha=1}^{M}B_{\nu_{\alpha}} \vert \Uparrow \rangle=\prod_{\alpha=1}^{M}\sum_{j} \frac{s_{j}^{-}}{\nu_{\alpha}-\epsilon_{j}}  \vert \Uparrow \rangle  \nonumber \\
=\sum\limits_{ j_{1}< \cdots <j_{M}}   \left[ \sum_{\mathcal{Q}}\frac{1}{\prod \limits_{\alpha=1}^{M}(\nu_{\alpha,k}-\epsilon_{\mathcal{Q}_{\alpha}})} \right] \vert   j_{1},\cdots,j_{M}  \rangle,
 \label{nu_ket}
\end{align}
where, as in the main text, $\mathcal{Q}$ are the permutations of indexes $\lbrace j_{1},\cdots,j_{M} \rbrace$. Note also that any of the two  ``$j$'' indexes are necessarily unequal. The overlap of state $\vert a_{1},\cdots,a_{M} \rangle$ with a BA eigenstate $\vert \nu_{1,k},\cdots,\nu_{M,k} \rangle$ can be simply obtained from the above expression as:
\begin{align}\label{olap}
\langle a_{1},\cdots,a_{M} \vert \nu_{1,k},\cdots,\nu_{M,k} \rangle = \sum_{\mathcal{P}}\frac{1}{\prod \limits_{\alpha=1}^{M}(\nu_{\alpha,k}-\epsilon_{{\mathcal{P}_{\alpha}}})},
\end{align}
where $\mathcal{P}$ are the permutations of indexes $\lbrace a_{1},\cdots, a_{M} \rbrace$. The above formula can be also written  in terms of  the permanent of the matrix $M_{kj}=1/(\nu_{\alpha,k}-\epsilon_{a_j})$ \cite{AFJPA}.

The scalar product between eigenstates can be obtained by insertion of $\sum_{ j_{1},\cdots,j_{M}} 
 \vert   j_{1},\cdots,j_{M}  \rangle\langle j_{1},\cdots,j_{M} \vert $:
\begin{align}
\langle & \nu_{M,k},\cdots,\nu_{1,k}\vert \nu_{1,k^{\prime}},\cdots,\nu_{M,k^{\prime}} \rangle = \frac{\delta_{kk^{\prime}}}{\vert N_{\nu_{k}}\vert^{2}} \nonumber \\
=&\sum\limits_{ j_{1}< \cdots <j_{M}}  \left[ \sum_{\mathcal{Q}}\frac{1}{\prod \limits_{\alpha=1}^{M}(\nu_{\alpha,k}-\epsilon_{{\mathcal{Q}_{\alpha}}})} \right] \nonumber \\
& \times\left[ \sum_{\mathcal{Q}}\frac{1}{\prod \limits_{\alpha=1}^{M}(\nu_{\alpha,k^{\prime}}-\epsilon_{{\mathcal{Q}_{\alpha}}})} \right],
\label{Eqkk}
\end{align}
where we used the overlaps Eq.~(\ref{olap}) between the two basis sets. The above expression gives the normalization coefficient $\vert N_{\nu_{k}}\vert^{2}$.
   
We can also derive another useful formula: 
\begin{align}
\langle j_{1} ,\cdots, j_{M}\vert  a_{1},\cdots, a_{M} \rangle = \sum_{\mathcal{Q}\in \lbrace j_{1},\cdots,j_{M} \rbrace}\prod_{\alpha=1}^{M} \delta_{a_{\alpha},j_{\mathcal{Q}_{\alpha}}} \nonumber \\
=\sum\limits_k  \sum_{\mathcal{Q} }\sum_{\mathcal{P} } \frac{\vert N_{\nu_{k}}\vert^{2}}{\prod \limits_{\alpha=1}^{M}(\nu_{\alpha,k}-\epsilon_{{\mathcal{Q}_{\alpha}}})\prod \limits_{\alpha=1}^{M}(\nu_{\alpha,k}-\epsilon_{{\mathcal{P}_{\alpha}}})}  
\label{Eqja}
\end{align}
which is obtained by inserting the resolution of the identity in terms of the BA eigenstates.

\section{Evaluation of physical observables}\label{appendix_sz}

We now discuss the explicit evaluation of Eq.~(\ref{Szt}), where we recall that $\vert \phi_{k} \rangle=N_{\nu_{k} }\vert \nu_{1,k},\cdots,\nu_{M,k} \rangle$ are the normalized eigenfunctions. Therefore, for an initial state in the form $\vert \Phi_{0} \rangle=s_{a_{1}}^{-} \cdots s_{a_{M}}^{-} \vert \Uparrow \rangle=\vert a_{1},\cdots,a_{M}\rangle$, the overlap matrix elements $\langle \Phi_{0} \vert \phi_{k} \rangle$ are immediately found from Eq.~(\ref{olap}), and are given in Eq.~(\ref{Phi_0}). Instead, to compute the matrix element of $s_{j}^{z} $, we use the fact that such operator is diagonal in the natural basis $\vert j_{1},\cdots,j_{M} \rangle $ and: 
\begin{align}
\langle j_{1},\cdots,j_{M} \vert  s_{j}^{z}  \vert   j_{1},\cdots,j _{M}  \rangle  = \frac{1}{2}-\sum_{\alpha=1} \delta_{jj_{\alpha}}.
\end{align}
By inserting the completeness relation in terms of the $\vert j_{1},\cdots,j_{M} \rangle $ states, and computing the overlaps with eigenstates $\vert \phi_{k} \rangle$ as in Eq.~(\ref{Phi_0}), it is easy to find that:
\begin{align}\label{sz_matrix_element}
\langle  \phi_{k}  \vert s_{j}^{z} \vert \phi_{k^{\prime}}  \rangle
= N_{\nu_{k} }^{*} N_{\nu_{k^{\prime}} } \sum\limits_{ j_{1}< \cdots <j_{M}} \left(\frac{1}{2}-\sum_{\alpha=1}^{M}\delta_{j j_{\alpha}}\right) \nonumber \\
  \times \left[ \sum\limits_{\mathcal{Q}}\frac{1}{\prod \limits_{\alpha=1}^{M}(\nu_{\alpha,k}-\epsilon_{{\mathcal{Q}_{\alpha}}})} \right] 
 \left[ \sum\limits_{\mathcal{Q}}\frac{1}{\prod \limits_{\alpha=1}^{M}(\nu_{\alpha,k^{\prime}}-\epsilon_{{\mathcal{Q}_{\alpha}}})} \right].
\end{align}
 Finally, combining overlaps and matrix element gives:
\begin{align}\label{sz_appendix}
s_{j}^{z}(t)=&\sum_{k,k'}  \cos(w_{kk^{\prime}}t) \sum\limits_{ j_{1}< \cdots <j_{M}} \left(\frac{1}{2}-\sum_{\alpha=1}^{M}\delta_{j j_{\alpha}}\right) \nonumber \\
&\times \left[ \sum\limits_{\mathcal{P}}\sum\limits_{\mathcal{Q}}\frac{\vert N_{\nu_{k}} \vert^2}{\prod \limits_{\alpha=1}^{M}(\nu_{\alpha,k}-\epsilon_{{\mathcal{P}_{\alpha}}})\prod \limits_{\alpha=1}^{M}(\nu_{\alpha,k}-\epsilon_{{\mathcal{Q}_{\alpha}}})} \right]\nonumber \\
&\times \left[ \sum\limits_{\mathcal{P}}\sum\limits_{\mathcal{Q}}\frac{\vert N_{\nu_{k'}} \vert^2}{\prod \limits_{\alpha=1}^{M}(\nu_{\alpha,k'}-\epsilon_{{\mathcal{P}_{\alpha}}})\prod \limits_{\alpha=1}^{M}(\nu_{\alpha,k'}-\epsilon_{{\mathcal{Q}_{\alpha}}})} \right].
\end{align}
In the above formulas, as usual, $\mathcal{P}$ are permutations of the indexes $ \lbrace a_{1},\cdots,a_{M} \rbrace$ (from the initial state) and $\mathcal{Q}$ are permutations of the indexes $ \lbrace j_{1},\cdots,j_{M} \rbrace$. The term proportional to 1/2 (from the parenthesis in the first line) can be further simplified as follows. First we note that this term contains a factor of type Eq.~(\ref{Eqkk}), which is $\propto \delta_{k k'}$. Therefore, we can simply set $\cos w_{kk'}t=1$ in this particular contribution. After noticing this, we can use Eq.~(\ref{Eqja}) to perform the summations over $k$ and $k'$ and finally recover the 1/2 appearing in Eq.~(\ref{sjz}) of the main text. 

Using a similar method, we can derive the spin-spin correlation function, given by 
 \begin{equation}
G^{z}_{mn}(t)=\sum_{k} \sum_{k^{\prime}} \langle \Phi_{0} \vert \phi_{k} \rangle  \langle  \phi_{k}  \vert s_{m}^{z} s_{n}^{z}\vert \phi_{k^{\prime}}  \rangle \langle  \phi_{k^{\prime}}  \vert \Phi_{0} \rangle e^{-i w_{kk^{\prime}}t}.
\end{equation}
While the overlaps are found in Eq.~(\ref{Phi_0}), the  matrix elements of $s_{m}^{z} s_{n}^{z}$ can be obtained in a way completely analogous to Eq.~(\ref{sz_matrix_element}):
\begin{align}\label{szz_matrix_element}
& \langle   \phi_{k}  \vert s_{m}^{z} s_{n}^{z} \vert \phi_{k^{\prime}}  \rangle
\nonumber \\ 
& = N_{\nu_{k} }^{*} N_{\nu_{k^{\prime}} } \sum\limits_{ j_{1}< \cdots <j_{M}} 
\left(\frac{1}{2}-\sum_{\alpha=1}^{M}\delta_{m j_{\alpha}}\right)  
\left(\frac{1}{2}-\sum_{\alpha=1}^{M}\delta_{n j_{\alpha}}\right) \nonumber \\
&\times \left[ \sum\limits_{\mathcal{Q}}\frac{1}{\prod \limits_{\alpha=1}^{M}(\nu_{\alpha,k}-\epsilon_{{\mathcal{Q}_{\alpha}}})} \right] 
 \left[ \sum\limits_{\mathcal{Q}}\frac{ 1 }{\prod \limits_{\alpha=1}^{M}(\nu_{\alpha,k^{\prime}}-\epsilon_{{\mathcal{Q}_{\alpha}}})} \right],
\end{align}
leading to Eq.~(\ref{Eszsz}) in the main text.

Finally, we discuss how to recover the initial conditions, and prove the conservation of magnetization. When time $t \rightarrow0$, the spin distribution becomes
\begin{align}
s_{j}^{z}(0)=&\frac{1}{2}-\sum_{k}\sum_{k^{\prime}}  \sum\limits_{ j_{1}< \cdots <j_{M}} \sum_{\alpha=1}^{M}\delta_{jj_{\alpha}}\nonumber \\
&\times  \sum_{\mathcal{P}}\sum_{\mathcal{Q}} \frac{\vert N_{\nu_{k}}\vert^{2}}{\prod \limits_{\alpha=1}^{M}(\nu_{\alpha,k}-\epsilon_{{\mathcal{P}_{\alpha}}}) (\nu_{\alpha,k}-\epsilon_{{\mathcal{Q}_{\alpha}}})} \nonumber \\
&\times  \sum_{\mathcal{P}}\sum_{\mathcal{Q}} \frac{\vert N_{\nu_{k'}}\vert^{2}}{\prod \limits_{\alpha=1}^{M}(\nu_{\alpha,k'}-\epsilon_{{\mathcal{P}_{\alpha}}})(\nu_{\alpha,k'}-\epsilon_{{\mathcal{Q}_{\alpha}}})} ,
\end{align}
which is in a form suitable to Eq.~(\ref{Eqja}). Performing the summations over $k$ and $k^{\prime}$ we get:
\begin{align}
 s_{j}^{z}(0)&=\frac{1}{2}-\sum\limits_{ j_{1}< \cdots <j_{M}}  \left( \sum_{\alpha}^{M}\delta_{jj_{\alpha}} \right) \left[ \sum_{\mathcal{Q}}{\prod_{\alpha}^{M} \delta_{a_{\alpha},j_{\mathcal{Q_\alpha}}}} \right] \nonumber \\
&=\frac{1}{2}-\sum_{\alpha}^{M}\delta_{j,a_{\alpha}},
\end{align}
which  recovers the  initial spin distribution. For the total magnetization, computing $m^{z}=\sum_{j}s_{j}^{z}(t) $ leads to  $\sum \limits_{j,\alpha}\delta_{jj_{\alpha}}=M$ in the second term of Eq.~(\ref{Szt}). Finally, the summation over $k,k'$ can be computed as discussed after Eq.~(\ref{sz_appendix}), leading to the expected result $m^{z}=(N+1)/2-M$.

\begin{figure}
\begin{center}
\includegraphics[width=0.45\textwidth]{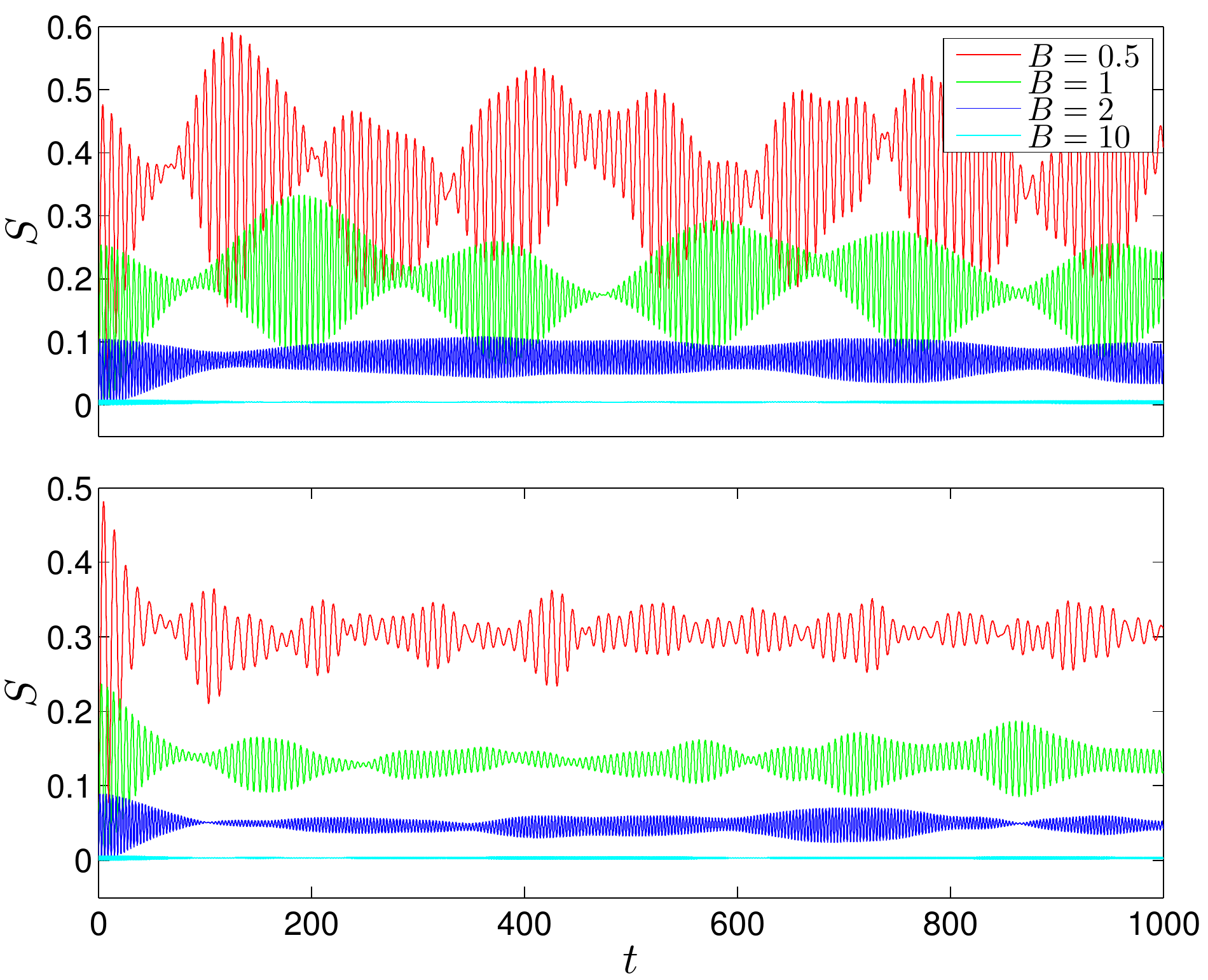}  
\end{center} 
\caption{Time dependence of the von Neumann entropy at different magnetic fields. upper subplot: initial state $\vert \Phi_{A} \rangle=\vert 0,7,8,9,10\rangle$. lower subplot: initial state $\vert \Phi_{B} \rangle=\vert 0,2,4,6,8\rangle$.}
\label{fig:VNE}
\end{figure}

\section{Reduced density matrix of the central spin}\label{sec:VNE}

The density matrix $\rho_{s}=\vert \psi(t) \rangle \langle\psi(t) \vert$  is given by 
\begin{align}
 \rho_{s} = &\sum_{k,k^{\prime}}  \vert \nu_{1,k},\cdots,\nu_{M,k} \rangle \langle \nu_{1,k^{\prime}},\cdots,\nu_{M,k^{\prime}} \vert  e^{iw_{kk^{\prime}}t}\nonumber \\
 & \times \left[ \sum_{\mathcal{P}}\frac{\vert N_{\nu_{k}}\vert^{2}}{\prod \limits_{\alpha=1}^{M}(\nu_{\alpha,k}-\epsilon_{{\mathcal{P}_{\alpha}}})} \right] 
\left[ \sum_{\mathcal{P}}\frac{\vert N_{\nu_{k^{\prime}}}\vert^{2}}{\prod \limits_{\alpha=1}^{M}(\nu_{\alpha,k^{\prime}}-\epsilon_{{\mathcal{P}_{\alpha}}})} \right] .
\end{align}
where, by using Eq.~(\ref{nu_ket}):
\begin{align}\label{rho_nu}
\vert & \nu_{1,k},  \cdots,\nu_{M,k} \rangle \langle \nu_{1,k^{\prime}},\cdots,\nu_{M,k^{\prime}} \vert \nonumber \\
&=\sum\limits_{ j_{1}< \cdots <j_{M}} \sum\limits_{ l_{1}< \cdots <l_{M}} \vert   j_{1},\cdots,j_{M}  \rangle  \langle    l_{1},\cdots,l_{M} \vert  \nonumber \\
& \times   \left[ \sum_{\mathcal{Q}}\frac{1}{\prod \limits_{\alpha=1}^{M}(\nu_{\alpha,k}-\epsilon_{{\mathcal{Q}_{\alpha}}})} \right] 
\left[ \sum_{\mathcal{R}}\frac{1}{\prod \limits_{\alpha=1}^{M}(\nu_{\alpha,k^{\prime}}-\epsilon_{{\mathcal{R}_{\alpha}}})} \right].
\end{align}
Here $\mathcal{R}$ indicates the permutations of $\lbrace l_{1},\cdots,l_{M} \rbrace$.

Now we consider the reduced density matrix of central spin, defined by $\rho_{cs}=\Tr_{1\rightarrow N}[\rho_{s}]$. With our choice of initial state, there are no off-diagonal elements and the result is in the form:
\begin{equation}
\rho_{cs}=A(t)\vert\uparrow\rangle \langle\uparrow \vert+D(t)\vert\downarrow\rangle \langle\downarrow \vert.
\end{equation}
The first contribution arises from the terms of Eq.~(\ref{rho_nu}) where $j_i=l_i > 0$ (i.e., site $0$ is not flipped) and $A(t)$ has the following expression:
\begin{align}
A(t)& =  \sum_{k,k^{\prime}} \sum_{0< j_{1}<\cdots<j_{M}} \cos(w_{kk^{\prime}}t) \nonumber \\
&\times \left[ \sum_{\mathcal{P}}\frac{\vert N_{\nu_{k}}\vert^{2}}{\prod \limits_{\alpha=1}^{M}(\nu_{\alpha,k}-\epsilon_{{\mathcal{P}_{\alpha}}})} \right] 
\left[ \sum_{\mathcal{P}}\frac{\vert N_{\nu_{k^{\prime}}}\vert^{2}}{\prod \limits_{\alpha=1}^{M}(\nu_{\alpha,k^{\prime}}-\epsilon_{{\mathcal{P}_{\alpha}}})} \right] \nonumber \\
&\times  \left[ \sum_{\mathcal{Q}}\frac{1}{\prod \limits_{\alpha=1}^{M}(\nu_{\alpha,k}-\epsilon_{\mathcal{Q}_{\alpha}})} \right]  
\left[ \sum_{\mathcal{Q}}\frac{1}{\prod \limits_{\alpha=1}^{M}(\nu_{\alpha,k^{\prime}}-\epsilon_{\mathcal{Q}_{\alpha}})} \right].
\end{align}
The second contribution corresponds to $j_1=l_1 =0 $ (i.e., the central spin is flipped) and $D(t)$ reads:
\begin{align}
D(t)& =  \sum_{k,k^{\prime}} \sum_{j_{1}=0 <j_{2}\cdots<j_{M}} \cos(w_{kk^{\prime}}t) \nonumber \\
&\times \left[ \sum_{\mathcal{P}}\frac{\vert N_{\nu_{k}}\vert^{2}}{\prod \limits_{\alpha=1}^{M}(\nu_{\alpha,k}-\epsilon_{{\mathcal{P}_{\alpha}}})} \right] 
\left[ \sum_{\mathcal{P}}\frac{\vert N_{\nu_{k^{\prime}}}\vert^{2}}{\prod \limits_{\alpha=1}^{M}(\nu_{\alpha,k^{\prime}}-\epsilon_{{\mathcal{P}_{\alpha}}})} \right] \nonumber \\
&\times  \left[ \sum_{\mathcal{Q}}\frac{1}{\prod \limits_{\alpha=1}^{M}(\nu_{\alpha,k}-\epsilon_{\mathcal{Q}_{\alpha}})} \right]  
\left[ \sum_{\mathcal{Q}}\frac{1}{\prod \limits_{\alpha=1}^{M}(\nu_{\alpha,k^{\prime}}-\epsilon_{\mathcal{Q}_{\alpha}})} \right],
\end{align} 
With the help of Eqs.~(\ref{Eqkk}) and (\ref{Eqja}), we can prove $A+D=1$. The von Neumann entropy is simply given as $S=-A\ln A-D\ln D$, and a representative plot is shown in Fig.~\ref{fig:VNE}.

\section{Numerical method for solving BA roots}
\label{sec:BAroot}
In our work we solve the BA equations (\ref{BAE}) using the numerical method presented in \cite{AFprb}. For convenience of the reader, in this Appendix we present the numerical method in detail. For general $M$, the asymptotic solution to Eq.~(\ref{BAE}) with $g \rightarrow 0$ reads:
\begin{equation}
\nu_{\alpha}(0)\approx \epsilon_{j}+\frac{g}{2}+\mathcal{O}(g^2).
\end{equation}
We then introduce the quantities
\begin{equation}
\Lambda_{j}=\sum_{\alpha=1}^{M} \frac{1}{\nu_{\alpha}-\epsilon_{j}},
\label{eqlamd}
\end{equation}
and transform Eq.(\ref{BAE}) to the following form:
\begin{equation}
\Lambda_{j}^{2}=-\sum_{l \neq j} \frac{\Lambda_{j}-\Lambda_{l}}{\epsilon_{j}-\epsilon_{l}}+\frac{2}{g}\Lambda_{j}.
\label{BAlamd}
\end{equation}
 The advantage of using the $\Lambda_j$ variables is in avoiding the singular dependence of $\nu_\alpha$ on system parameters. With $\nu_{\alpha}(0)$ as initial value, we use Newtonian iteration to find the solution of the above equations at fixed $g$. 

\begin{figure}
\begin{center}
\includegraphics[width=0.45\textwidth]{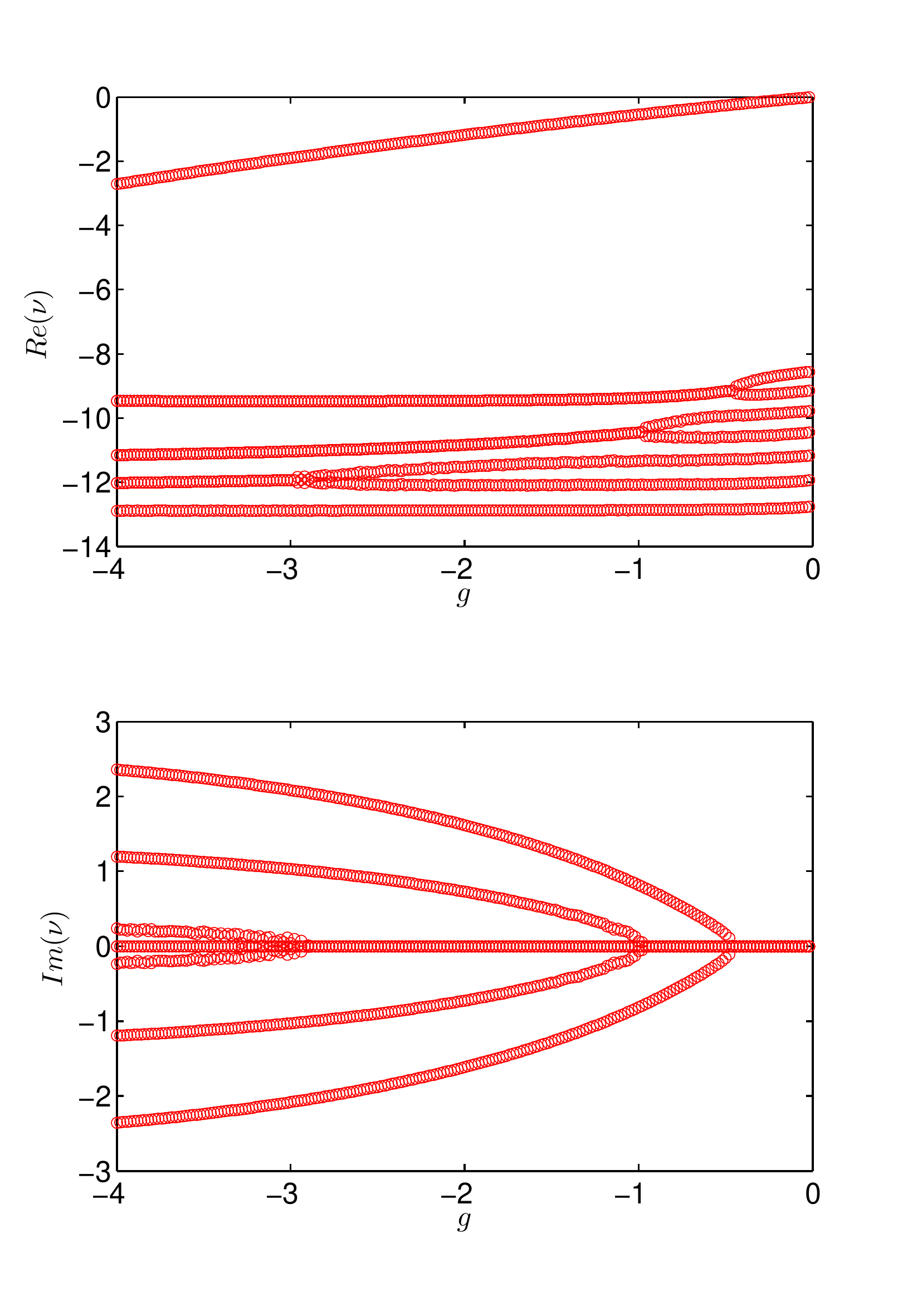}  
\end{center} 
\caption{Dependence of BA parameters on $g$. The top (bottom) panel shows thir real (imaginary) part. System parameters: $N=15$, $M=8$, and $A=2$.}
\label{fig:BAR}
\end{figure} 

After obtaining $\Lambda_{j}$, the $\nu_j$ can be found as follows. We first transform Eq.~(\ref{eqlamd}) into
\begin{equation}\label{pi_system}
\Lambda_{j} \prod_{\alpha=1}^{M}(\nu_{\alpha}-\epsilon_{j}) =\sum_{\alpha=1}^{M} \prod_{\beta \neq \alpha}(\nu_{\beta}-\epsilon_{j}).
\end{equation}
The above Eq.~(\ref{pi_system}) gives a linear system in the $p_i$ variables:
\begin{align}
p_{1}&=\sum_{\alpha} \nu_{\alpha} ,\nonumber \\
p_{2}&=\sum_{\alpha<\beta} \nu_{\alpha}\nu_{\beta} ,\nonumber \\
 &\hspace{2em } \ldots \nonumber \\
p_{M}&=\sum_{\alpha_{1}<...<\alpha_{M}} \nu_{\alpha_{1}} ... \nu_{\alpha_{M}} , 
\end{align}
which are elementary symmetric polynomials of the BA parameters $\nu_{\alpha}$ and $p_{0}$ can be defined as $1$.
Finally, after obtaining the values of $p_i$ we consider the $M$th-order polynomial:
\begin{equation}
F(\mu)=\prod_{\alpha}^{M}(\mu-\nu_{\alpha})=p_{0}\mu^{M}-p_{1}\mu^{M-1}+\ldots (-1)^{M} p_{M},  
\end{equation}
whose zeroes give the desired BA parameters $\nu_{\alpha}$.
An example of the dependence of the BA parameters $\left\lbrace \nu \right\rbrace$ on $g$ is shown in Fig.~\ref{fig:BAR}. Close to $g=0$ the BA parameters $\nu$ are all real, but complex conjugate pairs are formed by neighboring roots at larger values of $|g|$. The interested reader can also see \cite{AFprb} for the description of this method.

\end{appendix}



\begin{thebibliography}{99}
\bibitem{SuppM} See the appendix for details. There we present brief derivations for the time evolution of the spin distribution function, the spin-spin correlation function, and the Loschmidt echo for the Gaudin magnet with arbitrary number of flipped spins $M$.
\end{thebibliography}
\end{document}